\documentclass[aps,reprint,twocolumn,superscriptaddress]{revtex4-1}

\usepackage{soul}
\usepackage{xcolor}
\usepackage{amsmath}
\usepackage{graphicx}
\usepackage{epstopdf}
\usepackage{mathtools}
\usepackage{upgreek}
\usepackage{notes2bib}

\begin{document}
	\title{Hierarchical construction of higher-order exceptional points}
	%\title{Higher order exceptional points with supersymmetry between real and synthetic dimensions}
	
	\author{Q. Zhong}
		\affiliation{Department of Physics, Michigan Technological University, Houghton, Michigan 49931, USA}
		\affiliation{Henes Center for Quantum Phenomena, Michigan Technological University, Houghton, Michigan 49931, USA}
		
    \author{J. Kou}
    \affiliation{Division of Engineering and Applied Science, California Institute of Technology, Pasadena, California 91125, USA}
	
	\author{\c{S}. K. {\"O}zdemir }
	\affiliation{Department of Engineering Science and Mechanics, and Materials Research Institute, The Pennsylvania State University, University Park, Pennsylvania 16802, USA}

	\author{R. El-Ganainy}
	\email[Corresponding author: ]{ganainy@mtu.edu}
	\affiliation{Department of Physics, Michigan Technological University, Houghton, Michigan 49931, USA}
	\affiliation{Henes Center for Quantum Phenomena, Michigan Technological University, Houghton, Michigan 49931, USA}
	%\affiliation{Department of Electrical and Computer Engineering, Michigan Technological University, Houghton, Michigan, 49931, USA}

\begin{abstract}
	The realization of higher-order exceptional points (HOEPs) can lead to orders of magnitude enhancement in light-matter interactions beyond the current fundamental limits. Unfortunately,  implementing HOEPs in the existing schemes is a rather difficult task, due to the complexity and sensitivity to fabrication imperfections. Here we introduce a hierarchical approach for engineering photonic structures having HOEPs that are easier to build and more resilient to experimental uncertainties. We demonstrate our technique by an example that involves parity-time (PT) symmetric optical microring resonators with chiral coupling among the internal optical modes of each resonator. Interestingly, we find that the uniform coupling profile is not required to achieve HOEPs in this system --- a feature that implies the emergence of HOEPs from disorder and	provides resilience against some fabrication errors. Our results are confirmed by using full-wave simulations based on Maxwell's equation in realistic optical material systems.  
\end{abstract}
\maketitle
\textbf{Introduction---} The notion of non-Hermitian photonics \cite{El-Ganainy2018NP,Feng2017NP,El-Ganainy2019CP,Ozdemir2019NM}, together with its central concept of exceptional points (EPs) \cite{,Miri2019S} have attracted considerable attention in the past few years.  At an EP, two or more eigenvalues and their corresponding eigenstates/eigenvectors coalesce. As a result, the dimensionality of the eigenspace is reduced. An interesting feature of systems operating at EPs is their strong (nonlinear) response to small perturbation \cite{Kato-PTLO,Edelman1998LAA}. More specifically, for an EP of order $N$ (or EPN for short) formed by the coalescence of $N$ different eigenvalues and associated eigenstates, the response scales as $\sqrt[N]{\epsilon}$, where $\epsilon$ is the perturbation strength. Clearly, for $N>1$, this response `amplifies' weak effects, which could be potentially useful for sensing applications \cite{Wiersig2014PRL,Wiersig2016PRA, Hodaei2017N,Chen2017N}, wireless energy transfer \cite{Assawaworrarit2017N} and enhancing light-matter interactions \cite{Pick2017OE}. However, current schemes for achieving this requires the implementation of complex networks and  tuning of considerable number of design parameters \cite{Teimourpour2014PRA,Teimourpour-HOEP,Peng2014NP,Hodaei2017N,Nada2017PRB}.

In this work, we introduce a general approach for constructing tight-binding networks supporting higher-order exceptional points (HOEPs) out of arrangements that support only lower-order EPs. As an example, we apply our scheme for constructing photonic networks with HOEPs based on judicious engineering of the gain/loss profile and coupling coefficients in both  real and synthetic dimensions. More specifically, this strategy allows us to start with a parity-time (PT) symmetric arrangement having an $N^\text{th}$ order EP and modify it to obtain a $2N^\text{th}$ order EP in a straightforward fashion. Importantly, the resultant configuration requires tuning a small number of design parameters than those of conventional PT symmetric arrays with the same EP order, and hence is more robust against fabrication errors and experimental uncertainties.

\begin{figure}[!t]
	\centering
	\includegraphics[width=3in]{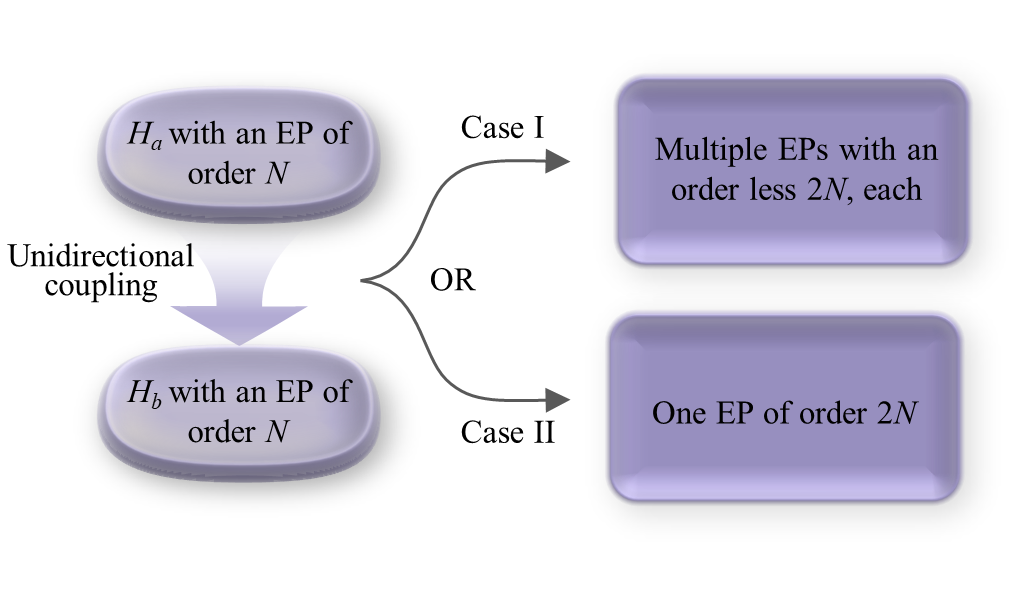}
	\caption{By introducing a unidirectional coupling between two different Hamiltonians $H_a$ and $H_b$, each of which exhibits an EP of order $N$, the combined system in general has $M$ EPs, each of which is of order $<2N$. As described in the text, under certain conditions $M=1$, the system has only one EP with order $2N$.}  
	\label{Fig.MainIdea} 
\end{figure}

\textbf{General procedure for doubling the order of EPs---} 
To this end, we start with two different discrete systems described by two matrix Hamiltonians $H_a$ and $H_b$, each with dimension $N \times N$ and exhibiting an EP of order $N$. Moreover, we assume that both EPs correspond to the same eigenvalue. We will denote the eigenvalue and eigenvector at the EPs by ($\mu$, $\vec{a}$) and ($\mu$, $\vec{b}$), respectively, where the $N$-dimensional  column vectors $\vec{a}$ and $\vec{b}$ are exceptional eigenvectors (i.e. the coalesced eigenvectors at the EP) of $H_a$ and $H_b$. Let us now consider the following mathematical construction:
\begin{equation} \label{Eq.H2N}
\mathcal{H}=
\begin{bmatrix}
H_a   &  \hat{0} \\
K    & H_b 
\end{bmatrix},
\end{equation}
where $K$ is an $N \times N$ matrix, and $\hat{0}$ is an $N \times N$ null matrix. Note that $\mathcal{H}$ has a dimension $2N \times 2N$. The eigenvalues and eigenvectors of $\mathcal{H}$ can be written as $\left \{ \lambda,\vec{c} \equiv \begin{bmatrix} \vec{c}_1 \\ \vec{c}_2 \end{bmatrix} \right \}$, where the superscript $T$ indicates matrix transpose; and $\vec{c}_{1,2}$ are each $N$-dimensional column vectors (thus the column vector $\vec{c}$ has a dimension of $2N$). The eigenvalue problem $\lambda I-\mathcal{H}$ can be then rearranged into:
\begin{subequations} \label{Eq.H2N_Eigen}
\begin{align}
(\lambda I-H_a) \vec{c}_1&= 0,  \label{Eq.H2N_Eigen_a} \\
(\lambda I-H_b) \vec{c}_2&= K \vec{c}_1. \label{Eq.H2N_Eigen_b}
\end{align}
\end{subequations}
Eq. (\ref{Eq.H2N_Eigen_a}) admits two possible solution: a nontrivial solution with an eigenvalue $\lambda=\mu$ and an eigenvector $\vec{c}_1=\vec{a}$; and a trivial one with $\vec{c}_1=\vec{0}$. Let us first consider the trivial solution. By substituting in Eq. (\ref{Eq.H2N_Eigen_b}), we obtain $(\lambda I-H_b) \vec{c}_2=0$, which has the solution $\lambda=\mu$ and $\vec{c}_2=\vec{b}$. On the other hand, if we consider the nontrivial solution of Eq. (\ref{Eq.H2N_Eigen_a}), we obtain  $(\mu I-H_b) \vec{c}_2=K \vec{a}$. Since $\text{rank}(\mu I-H_b)<N$, the last system of equations may be undetermined (i.e. it exhibits several solutions) or inconsistent (i.e. it has no solution), depending on the vector $K\vec{a}$. In the former case, the matrix $\mathcal{H}$ has, in general, $M$ eigenvectors that correspond to the eigenvalue $\mu$, where the actual value of $M$ depends on the system's details. The latter scenario is more interesting: it dictates that $\mathcal{H}$ has only one eigenvector $\vec{c}=\begin{bmatrix} \vec{0} \\ \vec{b}\end{bmatrix}$ corresponding to the eigenvalue $\mu$, i.e. the spectrum of $\mathcal{H}$ has an EP of order $2N$. These results are summarized in Fig. \ref{Fig.MainIdea}.

\begin{figure}[!t]
	\centering
	\includegraphics[width=3.4in]{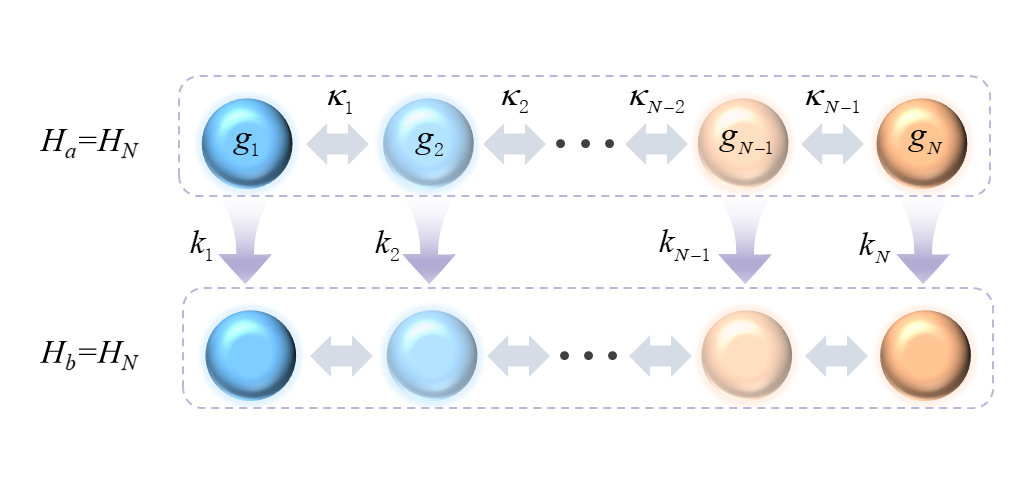}
	\caption{A schematic of two identical copies of a PT symmetric $J_x$ array. Each lattice has an EP of order $N$. By introducing disordered unidirectional coupling $k's$, the combined array can exhibit an EP of order $2N$.}
	\label{Fig.JX_array} 
\end{figure}

Having introduced the general framework, we now consider the following special case where: $H_a=H_b \equiv H$ are discrete PT symmetric Hamiltonians with an eigenvector $\vec{a}=\vec{b}=\vec{v}$, $K\equiv \text{diag}(k_1, k_ 2, ...,k_N)$ is a diagonal matrix whose elements represent a unidirectional coupling from the different elements of $H_a$ to the corresponding elements in $H_b$, as shown in Fig. \ref{Fig.JX_array}. Without any loss of generality, we also assume that $\mu=0$. Under these conditions, Eq. (\ref{Eq.H2N_Eigen_b})  reduces to $H \vec{c}_2=-K \vec{v}$. If this latter equation has one solution, then the spectrum of the combined system will contain two EPs of order $N$ each. On the other hand, if there is no solution, the spectrum will consist of only one EP of order $2N$. To investigate these two situations in more detail, we express $H$ in its Jordan canonical form as defined by the similarity transformation $H^J=S H S^{-1}$ with $S$ being the mapping matrix. The explicit form of $H^J$ is: 
\begin{equation} \label{Eq.HJ_def}
H^J =
\begin{bmatrix}
0   &  1  & 0 & \dots & 0 & 0 \\
0   &  0  & 1 & \dots & 0 & 0 \\
\vdots   &  \vdots  & \vdots   & \ddots & \vdots & \vdots \\
0   &  0  & 0 & \dots & 0 & 1 \\
0   &  0  & 0 & \dots & 0 & 0 \\
\end{bmatrix}, 
\vec{v}^J=
\begin{bmatrix}
1 \\
0 \\
\vdots \\
0 \\
\end{bmatrix}.
\end{equation}
In these bases $K^J=S K S^{-1}$, $\vec{c}_2^J=S\vec{c}_2$ and $\vec{v}^J=S\vec{v}$. Accordingly:
\begin{equation} \label{Eq.HJ_cond}
H^J\vec{c}_2^J =-K^J \vec{v}^J,\\
\end{equation}
If the diagonal matrix elements of $K$ are identical and equal to $k$, the combined structure  will respect PT symmetry, and we obtain $K^J=K=k I$ where $I$ is the $N \times N$ identity matrix. It follows that Eq. (\ref{Eq.HJ_cond}) is undetermined with a family of solutions given by $\vec{c}_2^J=[x,-k,0,...,0]^T$, where $x$ is a free parameter. Clearly, the spectrum of $\mathcal{H}$ in this case does not contain an EP of order $2N$. A more general argument applies when $[S,K]=0$. If, however, one introduces a disorder to the coupling matrix $K$ such that the $N^\text{th}$ element of the vector $K^J \vec{v}^J$ is not zero,  then Eq. (\ref{Eq.HJ_cond}) is inconsistent and has no solution, and  thus $\mathcal{H}$ would contain an EP of order $2N$.

\textit{Application to PT $J_x$ arrays---} In order to illustrate the power of the scheme presented above, we demonstrate its application to $J_x$ photonic array \cite{Christandl2004PRL, Perez-Leija2013PRA, Teimourpour2017PRApp} that posses PT symmetry \cite{Teimourpour2014PRA}. A tight binding network that realize an EP of order $N$ using PT $J_x$ arrays, will consist of $N$ sites and its Hamiltonian $H_{N}$ is given by \cite{Teimourpour2014PRA,Zhong2018PRA}:
\begin{equation}\label{Eq.HN}
H_N=
\begin{bmatrix}
i g_1 & \kappa_1 & \dots & 0 & 0 \\
 &  & \vdots &  &  \\
\dots& \kappa_{n-1} & i g_n & \kappa_n &\dots \\
 &   & \vdots &   &  \\
0 & 0 & \dots & \kappa_{N-1} & i g_N
\end{bmatrix},
\end{equation}
In the above equation, $g_n=(2n-N-1)\gamma$ and $\kappa_n=\sqrt{n (N-n)} \gamma$  with $n=1,2,...,N$, and $\gamma$ is a non-Hermitian parameter (in optics it represents gain or loss depending on its sign). Obviously, an implementation of an EP of order $2N$ can be directly obtained by scaling the above system according to $N \rightarrow 2N$ to obtain $H_{2N}$. This however introduces more non-uniformity in the coupling and gain/loss profiles, which poses practical limitations on realizing these arrangements. On the other hand, our scheme relaxes some of these constraints by enabling the construction of an EP of order $2N$ out of two exact copies of $H_N$ according to Eq. (\ref{Eq.H2N}) by substituting $H_a=H_b=H_N$, as demonstrated schematically in Fig. \ref{Fig.JX_array}. In what follows, we  take $K$ to a be a diagonal matrix and we will show that this choice can lead to an EP of order $2N$. Before we proceed, we highlight the interesting observation that both $H_{2N}$ and $\mathcal{H}_{2N}$ are connected via a similarity transformation. This can be demonstrated by noting that both $H_{2N}$ and $\mathcal{H}_{2N}$ have the same Jordan form which we denote by $H_{2N}^J$, i.e. $R^{-1}H_{2N}R=H_{2N}^J=S^{-1} \mathcal{H}_{2N} S$ for some mapping matrices $R$ and $S$. This connection can be also expressed in terms of discrete supersymmetry \cite{Miri2013PRL,El-Ganainy2015PRA}: by setting $A=RS$, $B=\mathcal{H}_{2N} (RS)^{-1}$, we find $H_{2N}=AB$ and $\mathcal{H}_{2N}=BA$.

Before we proceed further, we would like to mention that few studies have presented different routes for implementing higher order EPs. For instance, Nada et al. proposed a scheme that leads to high order EP without relying on gain/loss distribution \cite{Nada2017PRB}. Our work goes beyond previous studies by introducing a straightforward way of modifying an already implemented system that has $N$-th order EP to double the order of the EP to $2N$, with the minimal number of additional design parameters, and by showing for the first time that higher-order EPs can arise due to disorder, which completely goes against the conventional wisdom in the field.

\begin{figure}[!t]
	\centering
	\includegraphics[width=3.4in]{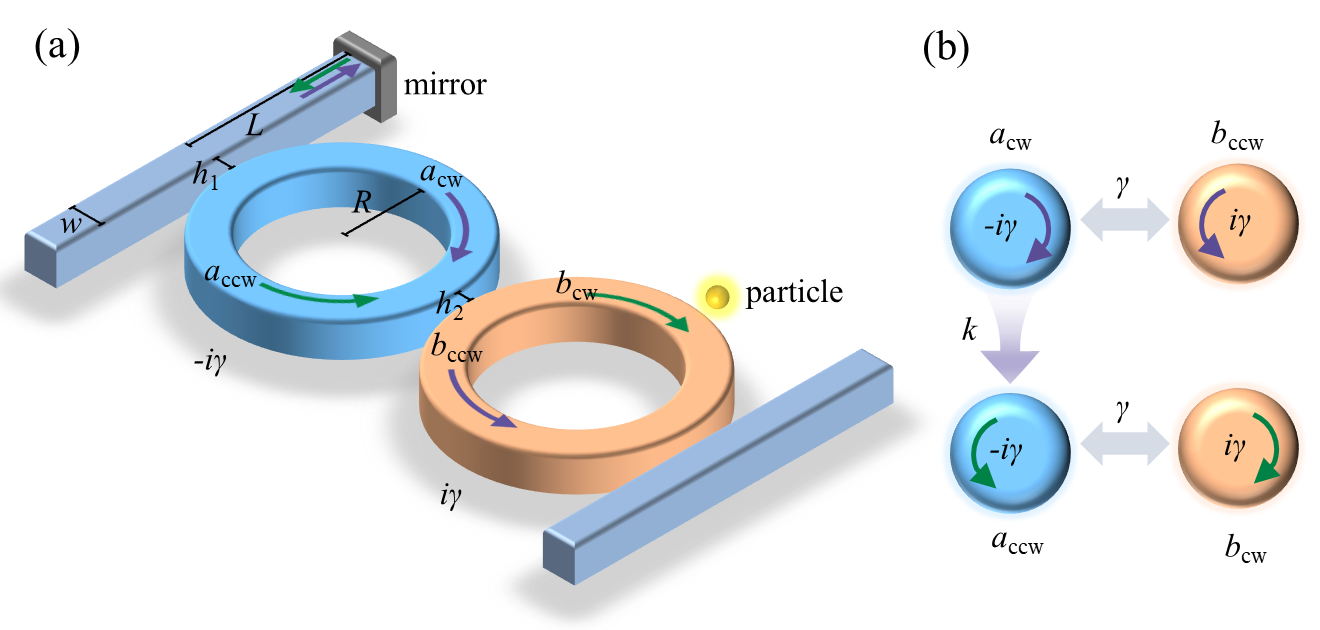}
	\caption{(a) A PT symmetric system consisting of two microring resonators having equal loss ($-i \gamma$), gain ($i \gamma$) and coupling to implement an EP of order two. By introducing an evanescent coupling between any of the rings (the lossy ring here) and a waveguide with an end mirror, the system realizes the Hamiltonian $\mathcal{H}_4$ of Eq. (\ref{Eq.H4}). The nanoparticle close to the gain ring is used as a small perturbation to confirm that the combined system indeed has an EP of order four in the later simulation. (b) Equivalent system in (a) in real space, detailing the loss/gain distribution and the coupling profile between the various CW/CCW components. }
	\label{Fig.Microrings} 
\end{figure}

\textbf{Implementation of $\mathcal{H}_{2N}$ using synthetic dimensions---}  In principle, the structure shown in Fig. \ref{Fig.JX_array} can be implemented by using $2N$ resonators  together with optical isolators to introduce the unidirectional coupling. This however does not provide any advantage in terms of scalability or fabrication. Alternatively, we explore a different route that relies on synthetic dimensions \cite{Celi2014PRL,Livi2016PRL,Ozawa2016PRA,Lustig2019N,Yuan2018O}. Particularly, we consider a PT symmetric $J_x$ array made of $N$  microresonators. Time reversal symmetry implies that each resonator supports two traveling waves at each resonant frequency, clockwise(CW) and counterclockwise (CCW). Taken together, these $2N$ modes can act as bases for implementing an EP of order $2N$. Figure \ref{Fig.Microrings}(a) illustrates one possible realization of the above photonic network when $N=2$. The unidirectional coupling here is introduced via an evanescently coupled waveguide with an end mirror --- a geometry that was recently proposed for building robust EP sensors \cite{Zhong2019PRL}, amplifiers \cite{Zhong2020PRApp}, and directional absorbers \cite{Zhong2019OL}. Figure \ref{Fig.Microrings}(b) depicts the equivalent arrangement in real space. Interestingly, if a mirror is introduced on each waveguide, the system will not exhibit a HOEP. In other words, the formation of the HOEP in this structure can be achieved only by breaking the PT symmetry -- a rather remarkable observation.

\textit{Applications in enhanced light-matter interaction---} Before we present a concrete design for the above structure with realistic material systems, we first confirm that it can be indeed used to enhance light-matter interaction. As a prototype example, we focus again on the implementation of $\mathcal{H}_4$ as shown in Fig. \ref{Fig.Microrings} and we assume that a nanoparticle exists in the near field region of one of the resonators as shown in Fig. \ref{Fig.Microrings}. By assuming that the nanoparticle introduces a perturbation strength $\epsilon$, it follows from the perturbation theory of non-Hermitian operators \cite{Kato-PTLO,Edelman1998LAA} that an eigenfrequency splitting scaling $\sqrt[4]{\epsilon}$ should take place. To confirm that this is indeed the case for our structure, we consider its Hamiltonian as derived by using temporal coupled mode theory \cite{Chen2018PR}, which takes the form $\mathcal{H}'_{4}=\mathcal{H}_4+H_p$, with:
\begin{equation}\label{Eq.H4}
\begin{aligned}
\mathcal{H}_{4}&=
\begin{bmatrix}
\omega_0-i \gamma & \gamma  & 0 & 0  \\
\gamma  & \omega_0+i\gamma & 0 & 0  \\
k  & 0 & \omega_0-i\gamma & \gamma \\
 0 & 0 & \gamma & \omega_0+i\gamma \\
\end{bmatrix}, \\
H_p&=  
\begin{bmatrix} 
0 & 0  & 0 & 0  \\
0 & \epsilon  & 0 & \epsilon  \\
0 & 0  & 0 & 0 \\
0 & \epsilon & 0 & \epsilon \\
\end{bmatrix}.
\end{aligned}
\end{equation}
Here $\omega_0$ is the resonant frequency of each mode, $\pm \gamma$ is the gain/loss coefficients of the modes in the gain and loss resonators, which is also taken to be identical to the coupling coefficient between the two resonators in order to implement a PT symmetric dimer (when $k=0$). Additionally,  $H_p$ is the perturbation Hamiltonian, and $\epsilon=|\epsilon|\exp{(i\phi_{\epsilon})}$. We remark that the above Hamiltonian is expressed in the bases $[a_\text{cw} , b_\text{ccw} , a_\text{ccw} , b_\text{cw}]^T$ (see Fig. \ref{Fig.Microrings}). When $\epsilon=0$, the above system has an EP4 with an eigenvalue $\omega=\omega_0$. Assuming a small perturbation $|\epsilon| \ll |\gamma|, |k|$, we obtain the following asymptotic expression (see Appendix A for more details) for the eigenvalues $\omega_n$: 
\begin{equation}\label{Eq.Splitting}
\omega_n-\omega_0 = \chi_n \sqrt[4]{|\epsilon|} + \mathcal{O}\left(\sqrt[4]{|\epsilon|^3}\right),
\end{equation}
where $\chi_n=i^{n-1} \sqrt[4]{\exp(i \phi_\epsilon)k \gamma^2}$,  and $n=1,2,3,4$. In other words, by just evanescently coupling the resonators to waveguides and terminating one waveguide with a mirror, the eigenvalue splitting changes from $\sqrt{|\epsilon|}$ to $\sqrt[4]{|\epsilon|}$. Similarly, one can apply the same strategy to the PT system that supports EP of order three, which has been implemented experimentally in \cite{Hodaei2017N}, in order to obtain a modified structure supporting an EP of order six.

\begin{figure}[!t]
	\centering
	\includegraphics[width=3.4in]{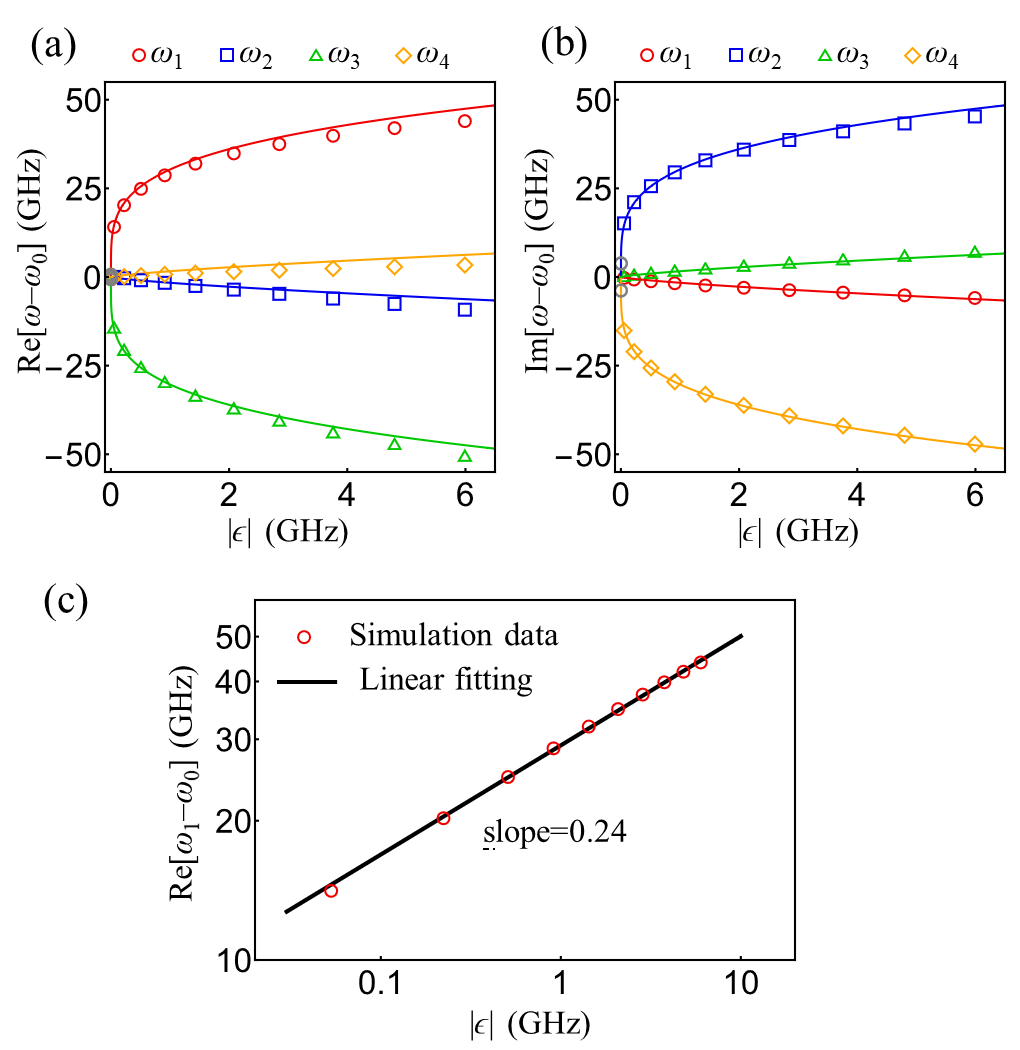}
	\caption{Real (a) and imaginary (b) splitting of the  eigenfrequency of the system in Fig. \ref{Fig.Microrings} as a result of a perturbation  $|\epsilon|$ (see Appendix C) introduced by a nanoparticle. Dots represent the results obtained from full-wave FEM simulations for the ring resonator parameters given in the Appendix C, while the solid black line is a plot of the analytical formula of Eq. (\ref{Eq.Splitting}).  (c) Log scale plot of the real frequency shift $\text{Re}[\omega_1-\omega_0]$ as a function of $|\epsilon|$. The slope of the linear fit is $\sim 0.24$, confirming that the system is operating in the vicinity of an EP of order four.}  
	\label{Fig.Splitting} 
\end{figure}

\textit{Full-wave simulations---} Finally, we confirm our results by performing full-wave analysis using finite element method (FEM) \bibnote{The simulations were performed using COMSOL software package} for the structure of Fig. \ref{Fig.Microrings}(a) with realistic dimensions and material systems as detailed in the Appendix B. Full account of numerically extracting the relevant optical parameters is presented in Appendix C.  Figures \ref{Fig.Splitting}(a) and (b) present the real and imaginary values of the four eigenfrequencies of the system as a function of the perturbation induced by the nanoparticles of different radii (see Appendix D for the relation between $|\epsilon|$ and nanoparticle radius $R_p$), where the solid line represents analytical results obtained from Eq. (\ref{Eq.Splitting}), while the dots represent actual values obtained from full-wave numerical simulations using FEM. As expected, the perturbation due to the particle causes the unperturbed eigenstate to split into four different eigenmodes. From the figures, we observe good agreement between theory and simulations. We note however that even in the absence of the particle, our simulations indicate a finite splitting between the eigenfrequencies. This is due to the fact that the waveguides themselves can introduce back-scattering between the modes. This means that our system operates in the vicinity, rather than exactly at, the EP. As we will see below, this does not have a significant impact on the ability of the structure to enhance light-matter interactions.

Figure \ref{Fig.Splitting}(c) plots $\text{Re}[\omega_1-\omega_0]$ as a function of $\epsilon$ using log scale. A curve fitting of the data produces the relation $\text{Re}[\omega_1-\omega_0]=A \epsilon^{0.24}$, corresponding to a line with a slope $0.24$ on the log scale, which is very close to the $0.25$ value expected for an ideal EP of order four. Moreover, a comparison between the numerical and analytical values for the splitting amplitudes (i.e. the value of the constant $A$ in the formula above and $\chi_1$ in Eq.(\ref{Eq.Splitting})) confirms the excellent agreement with only $4\%$ error. For completeness, we discuss the modal profiles associated with the resonant modes of the structure investigated above in Appendix E.

\textbf{Conclusions---} In summary, we have introduced a new, generic approach for constructing tight-binding Hamiltonians with HOEPs out of initial arrangements with lower-order EPs. As an illustrative example, we have demonstrated the detailed application of our scheme to PT symmetric $J_x$ arrays. By focusing on PT symmetric dimer and utilizing the concept of synthetic dimensions in multimodal systems, we have presented an elegant and more robust (compared to conventional configurations) implementation of optical ring dimer that exhibits a fourth-order EP. We should note here that many interesting features of EPs, such as a system's response to perturbations as in sensors, stem from the order of EPs: the higher the better, regardless of whether it is an odd or even order EP. Mathematically speaking, there is no bound on the order of EP that can be engineered using our approach. Limitations originate from the maturity and scalability of the fabrication process of the underlined physical platform. For example, in electronics where fabrication is highly scalable and fabrication errors are minimal, one can construct EP of any even order EP for telemetry and wireless energy transfer. Scalability in photonics has always been a critical problem despite decent progress in the past few years, and in general building photonic systems with HOEP is challenging (i.e., none beyond order 3 yet). Our approach provides a methodological approach and a realistic route to build photonic systems with HOEP, alleviating some of the difficulties encountered. This in turn may have an impact on building better sensing devices (though this topic is still subject to debate, see for instance \cite{Langbein2018PRA,Zhang2019PRL,Lau2018NC,Jan2020NC,Jan2020PRA}), as well as photonic components with enhanced nonlinear interactions.

\begin{acknowledgments}
R.E. acknowledges fruitful discussions with J. Wiersig. R.E. acknowledges support from ARO (Grant No. W911NF-17-1-0481), NSF (Grant No. ECCS 1807552), the Max Planck Institute for the Physics of Complex Systems, and the Henes Center for Quantum Phenomena at Michigan Technological University. S.K.O acknowledges support from ARO (Grant No. W911NF-18-1-0043), NSF (Grant No. ECCS 1807485), and AFOSR (Award no. FA9550-18-1-0235).
\end{acknowledgments}

\appendix

\section{Eigenvalues of Hamiltonian under perturbation}
Here we sketch the derivation of the perturbative expression of Eq. (7) in the main text. The eigenvalues of $\mathcal{H}'_{4}=\mathcal{H}_4+H_p-\omega_0 I$ are obtained by solving $|z I-\mathcal{H}'_{4}|=0$, which gives the characteristic equation:
\begin{equation} \label{Eq.Perturb}
z^4-2 \epsilon z^3-2i\gamma \epsilon z^2-k \gamma^2 \epsilon=0.
\end{equation}
We are interested here in the small $\epsilon$ values. We thus employ a perturbation analysis based on Newton-Puiseux series, i.e. we expand $z=c_1 \epsilon^{\frac{1}{4}}+c_2 \epsilon^{\frac{2}{4}}+c_3 \epsilon^{\frac{3}{4}}+...$. By substituting back in Eq. (\ref{Eq.Perturb}), and solving for the coefficients $c's$, we find $c_1=i^{n-1} k^{\frac{1}{4}} \gamma^{\frac{1}{2}}$, $c_2=0$, and $c_3=-\frac{1}{2} i^{-n} k^{-\frac{1}{4}}\gamma^{\frac{1}{2}}$.

\section{Geometric and material parameters}

For the fullwave simulations, we used the following parameters: each of the rings and the waveguides has a width $w=0.25 \ \mu \text{m}$, and refractive indices of 3.47 in a background refractive index $n_0=1.44$ (relevant to SiN on silica platforms); the radii of the rings are identical and taken to be $R=4.75 \ \mu \text{m}$ each; the edge-to-edge separation between each ring and its neighbor waveguide is $h_1=0.25 \ \mu \text{m}$ while that between the two rings is  $h_2=0.4 \ \mu \text{m}$; the distance between the center of the nanoparticle and the edge of the ring resonator is $0.1 \ \mu \text{m}$; and the distance between the ring-waveguide junction and the mirror is $L=5.11 \ \mu \text{m}$. Moreover, the mirror is implemented by using a thin layer of silver with a thickness of 100 nm, and the loss and gain are modeled by including an imaginary part to the refractive index of the two microrings, i.e. $n_{1,2}=3.47 \pm i n_i $. Finally, the refractive index of the nano-particle is $n_p=3.47$. In order to minimize the effect of the surface roughness along the ring resonator due to numerical discretization, we set up the average mesh size equal to $\lambda/50$  (here $\lambda=\lambda_f/n$ where $\lambda_f=1550 \text{ nm}$ is some free space reference wavelength and $n$ is the refractive index of the materials) and use a finer mesh size of 1 nm around the nanoparticle region.

\section{Optical parameters of the structure}

\begin{figure}[!t]
	\centering
	\includegraphics[width=3.4in]{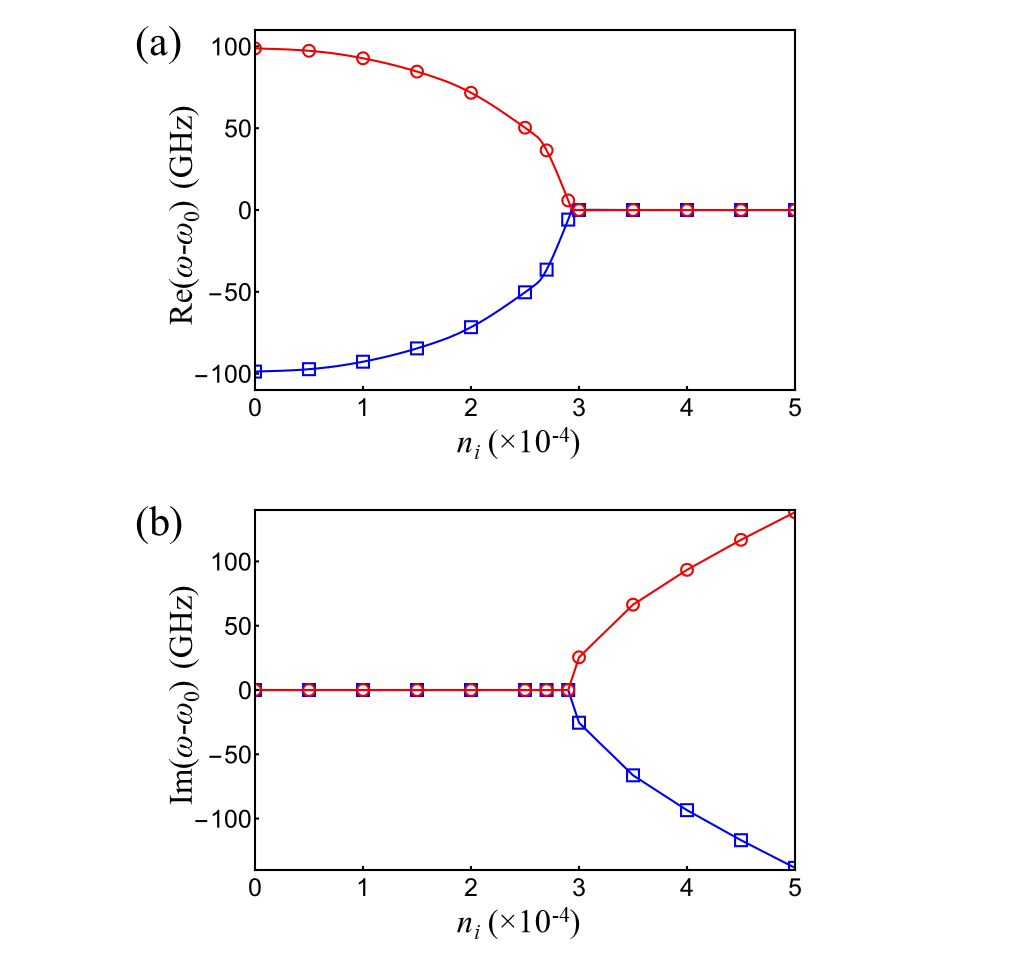}
	\caption{Real (a) and imaginary (b) parts of the resonant frequencies   $\omega$ of PT-symmetric resonators  as a function of the imaginary part of the optical index $n_i$ (see full description in the text) as obtained by full-wave simulations of the arrangement in Fig. 3  without the waveguide. The EP is located at $n_i=2.9051\times10^{-4}$. The coupling coefficient between the two resonators can be obtained from the data at $n_i=0$, which is found to be $\kappa=98.8\ \text{GHz}$.}  
	\label{Fig.EP_Test} 
\end{figure}

In order to extract the numerical values of the various optical parameters, we first simulate a single microring resonator and compute its resonance frequency $\omega_0$, or equivalently the corresponding free space wavelength of $1548\ \text{nm}$. 

\begin{figure}[!b]
	\centering
	\includegraphics[width=3.4in]{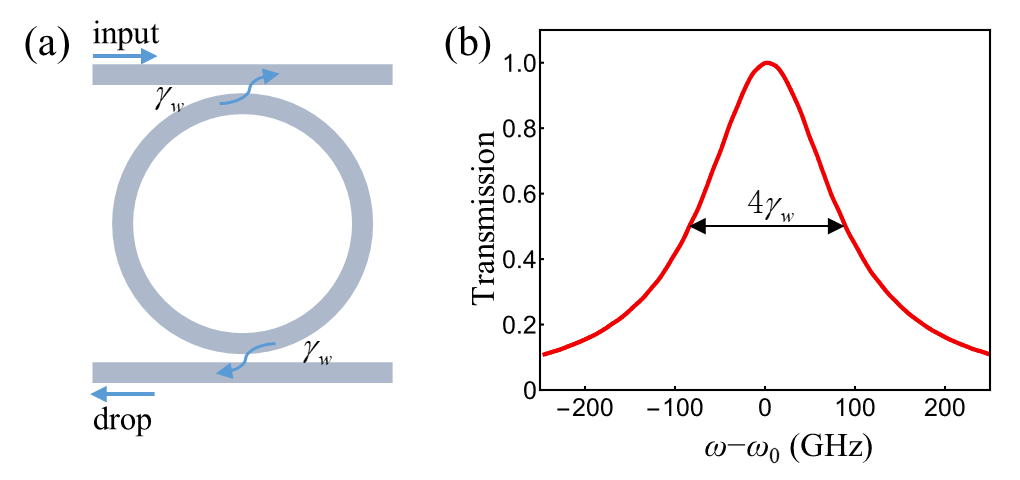}
	\caption{(a) A microring resonator in add-drop configuration is used to evaluate the decay rate from the resonator. (b) Power transmission spectrum as obtained by using full-wave simulations gives a full-width at half-maximum bandwidth of $4\gamma_w=173 \text{ GHz}$. }  
	\label{Fig.AddDrop} 
\end{figure}

Next, we consider a coupled resonators configuration without any waveguides. In this case, the coupling coefficient between the resonators can be obtained by calculating the resonant frequencies of the even/odd supermodes in the absence of any gain or loss. By doing so using COMSOL software package, we find $\kappa =98.8\ \text{GHz}$. Next, we include the gain and loss in the two resonators by adding an imaginary part $\pm n_i$ to their refractive indices in order to create a PT symmetric arrangement and we plot the resonant frequency as a function of $n_i$ as shown in Fig. \ref{Fig.EP_Test}. From these plots, we find that the EP occurs when $n_i=2.9051\times10^{-4}$, which indicate that the field amplitude gain and loss at this point is equal to the coupling between the resonators.

Finally we evaluate the decay rate from the microring to the  waveguide $\gamma_w$ by constructing a symmetric add-drop ring resonator device  and measuring the bandwidth of the transmission between the input and the output port as a function of frequency (Fig. \ref{Fig.AddDrop}(a)). In this case, for a negligible material absorption and radiation loss, the bandwidth is given by $BW=4\gamma_w$. The FEM simulations gives the value $\gamma_w=43.25 \ \text{GHz}$.  Thus, for an all-pass ring-waveguide structure, the resonant frequency is complex and is given by $\omega_0-i\gamma_w$.

\section{Perturbation coefficient $\epsilon$ due to the nano-particle}

\begin{figure}[!b]
	\centering
	\includegraphics[width=3.4in]{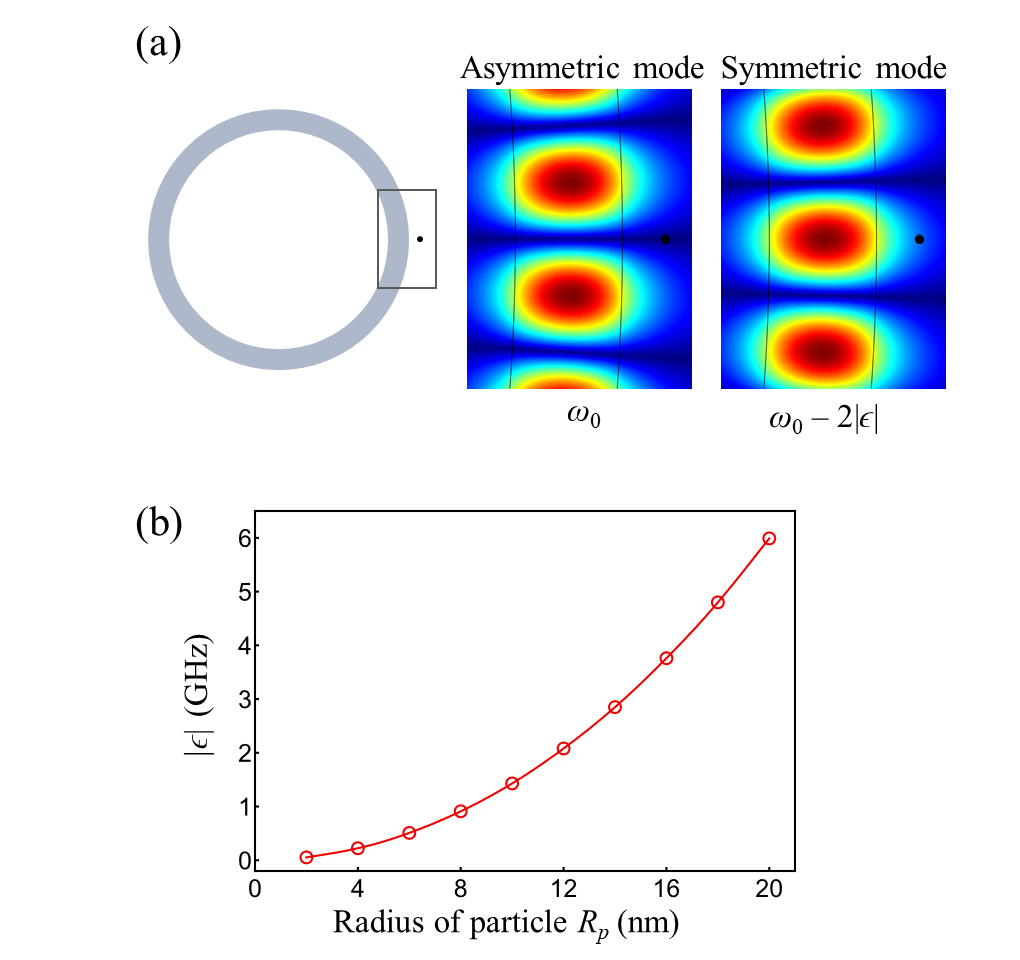}
	\caption{ A nanoparticle in the vicinity of a microring resonator (left panel of (a)) will introduce a coupling between the CW and CCW modes. The new supermodes of the ring will have symmetric/asymmetric field distribution with respect to the particle location. These modes are depicted in the right panel of (a) for a nanoparticle of radius 10-nm located at the node of the asymmetric mode and anti-node of the symmetric mode, respectively (the rest of the parameters as identical to those used in the text). (b) By using full-wave simulations to compute the frequency splitting due to nanoparticles and fitting these results to those obtained from the Hamiltonian $H_\text{test}$, we obtained the perturbation $|\epsilon|$ as a function of the particle's radii, $R_p$. (b) plots such a relation when $R_p$ varies from 2 to 20 nm.}  
	\label{Fig.Particle_Perturb} 
\end{figure}

Here we characterize the perturbation $\epsilon$ introduced by the nanoparticle as a function of its radius $R_p$. To do so, we first note that a particle located in the near field of the microresonators will introduce a coupling between the CW and CCW modes, forming standing wave patterns. These new supermodes are distributed such that the particle is located at the node of the antisymmetric mode, and at the maximum of the symmetric as shown in Fig. \ref{Fig.Particle_Perturb}(a) for a $10 \  \text{nm}$ particle. Here symmetric and antisymmetric refer to the field distribution with respect to a horizontal line that crosses the nano-particle's center. Consequently, the antisymmetric mode is not affected much by the particle while symmetric mode experiences a frequency shift of $2|\epsilon|$. The perturbation Hamiltonian $H_p$ in the main text, which is written in the CW/CCW bases, capture this behavior. Finally, Fig. \ref{Fig.Particle_Perturb}(b) plots the value of $|\epsilon|$ (the frequency is red-shifted because the particle's refractive index is larger than that of the surrounding \cite{Zhu2010NP}) as a function of $R_p$ when the latter varies from 2 to 20 nm. In obtaining this plot, we fit the results obtained from full-wave simulations to those obtained using the Hamiltonian $H_\text{test}=\begin{bmatrix}
\omega_0+\epsilon && \epsilon\\
\epsilon && \omega_0+\epsilon
\end{bmatrix}$.

\section{Modal profile}
\begin{figure*}[!t]
	\centering
	\includegraphics[width=5.4in]{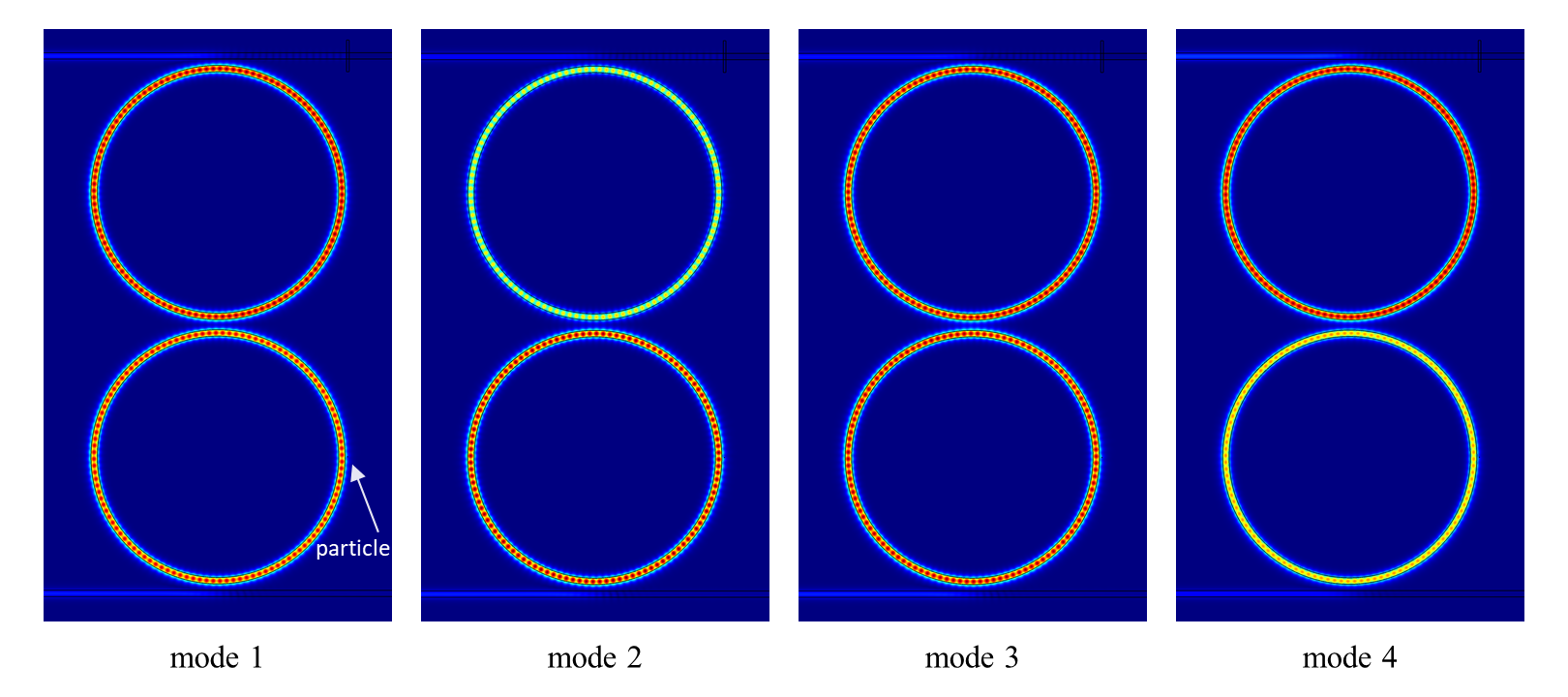}
	\caption{Amplitude of the electric field distribution associated with the four eigenmodes of the structure for a nanoparticle with radius 10 nm. Note that mode 1 and 3 have nearly identical intensity in the two microring, consistent with their dominant real frequency shift. On the other hand, mode 2 has more intensity in the gain microring (bottom one), while mode 4 has more intensity in the lossy microring (top one), and as a result, they experience strong imaginary frequency shifts.}  
	\label{Fig:Mode_Profiles} 
\end{figure*}

Figure \ref{Fig:Mode_Profiles} presents an example of the spatial field  distribution (amplitude of the electric field) of the four eigenmodes of the structure  for a particle of radius 10 nm.  We observe that both eigenmodes 1 and 3 have equal power distribution in both rings, which explain their dominant real frequency splitting as demonstrated in Fig. 4(a). On the other hand,  modes 2 and 4 have more intensity in the gain/loss microring, respectively which explain their dominant imaginary frequency shift as can be seen from Fig. 4(b). These numerical results thus confirm the analytical predictions and demonstrate the feasibility of our approach for practical implementations.

\newpage
\bibliographystyle{apsrev4-1}
\bibliography{Reference} 
\end{document}